\documentclass[11pt]{article}
\usepackage{graphicx,aas_macros}
\usepackage{amsmath,amssymb}
\title{Imaging Spectroscopy of Solar Radio Burst Fine Structures}

{\small \author{ E.~P.~Kontar$^{1}$,
S.~Yu$^{2,3}$, A.~A.~Kuznetsov$^{4}$,
A.~G.~Emslie$^{5}$, B.~Alcock$^{1}$, \\
N.~L.~S.~Jeffrey$^{1}$,
V.~N.~Melnik$^{6}$, N.~H.~Bian$^{1}$,  P.~Subramanian$^{7}$\\
$^{1}$ School of Physics and Astronomy, University of Glasgow, Glasgow G12 8QQ, UK\\
$^{2}$  New Jersey Institute of Technology, Newark, New Jersey 07102,  USA\\
$^{3}$ National Astronomical Observatories, Chinese Academy of Sciences, Beijing 100012, China\\
$^{4}$ Institute of Solar-Terrestrial Physics, Irkutsk 664033, Russia\\
$^{5}$ Department of Physics \& Astronomy, Western Kentucky University, Bowling Green, Kentucky 42101, USA\\
$^{6}$ Institute of Radio Astronomy, National Academy of Sciences of Ukraine, Kharkiv 61002, Ukraine\\
$^{7}$ Indian Institute of Science Education and Research, Pune 411008, India}}

\newcommand{\lapprox} {\, \lower3pt\hbox{$\sim$}\llap{\raise2pt\hbox{$<$}}\,}
\newcommand{\gapprox} {\, \lower3pt\hbox{$\sim$}\llap{\raise2pt\hbox{$>$}}\,}

\begin{document}

\maketitle

\begin{abstract}
Solar radio observations provide a unique diagnostic of the outer solar atmosphere. However, the inhomogeneous turbulent corona strongly affects the propagation of the emitted radio waves, so decoupling the intrinsic properties of the emitting source from the effects of radio-wave propagation has long been a major challenge in solar physics.  Here we report quantitative spatial and frequency characterization of solar radio burst fine structures observed with the LOw Frequency Array (LOFAR), an instrument with high time resolution that also permits imaging at scales much shorter than those corresponding to radio-wave propagation in the corona. The observations demonstrate that radio-wave propagation effects, and not the properties of the intrinsic emission source, dominate the observed spatial characteristics of radio burst images.  These results permit more accurate estimates of source brightness temperatures, and open opportunities for quantitative study of the mechanisms that create the turbulent coronal medium through which the emitted radiation propagates.
\end{abstract}

\section{Introduction}
During sporadic periods of activity, the Sun produces the largest magnetic energy release events in the solar system: solar flares and coronal mass ejections (CMEs). Flares emit radiation across the electromagnetic spectrum from gamma- and X-rays\cite{2011SSRv..159..107H} to radio waves\cite{2008A&ARv..16....1P}. Solar radio bursts originate from the acceleration of electrons in the relatively tenuous (electron number density $n \lapprox 10^8$~cm$^{-3}$) solar corona, a region that, because of its low plasma density, produces very low, and hence undetectable, levels of X-ray and Extreme Ultra-Violet (EUV) emission. The radio bursts produced in such regions are, however, easily observable and thus provide unique diagnostics of electron acceleration and propagation in the outer corona and surrounding heliosphere. They provide information on the impulsive initial evolution of solar eruptions, information that is essential to the overall understanding of such events and hence to developing an effective system of space weather prediction and mitigation.

Most of the brightest solar radio bursts are due
to coherent radio plasma emission processes\cite{1980panp.book.....M}: the injection of non-thermal electrons into the solar corona leads to the generation of Langmuir plasma waves through the electron-electron two-stream instability\cite{1980panp.book.....M}, and these plasma waves are converted into radio emission at both the plasma (fundamental) $f_{\rm pe} \simeq 9\times 10^{-3} \sqrt{n \mbox{ [cm$^{-3}$]}}$~MHz and second harmonic ($f_{\rm H} = 2 f_{\rm pe}$) frequencies. Radio bursts produced by electrons moving away from the Sun along open magnetic field lines are known as Type III bursts.  The fine frequency structures, so-called Type IIIb bursts,  are commonly believed to be caused by density inhomogeneities in the background plasma\cite{1975SoPh...40..421T,1982srs..work..182M}.
The radiation propagates through the solar corona where it is both refracted\cite{1971A&A....10..362S} and scattered by turbulent plasma processes\cite{1965BAN....18..111F,1974SoPh...35..153R},
hence understanding these propagation effects is critical to a correct interpretation of solar radio burst images\cite{1972PASAu...2..100S,1983PASAu...5..208R,1999A&A...351.1165A}. Because coherent plasma emission produces radio waves at frequencies close to the local plasma frequency\cite{1994ApJ...426..774B,1999A&A...351.1165A,2008ApJ...676.1338T}, propagation effects are particularly significant and must therefore be carefully considered in determining both the intrinsic properties of the surrounding plasma (e.g., density, magnetic field, density gradient, turbulence) and of the emitting high-energy electron beams (e.g., location, energy).

High-time-resolution one-dimensional scans\cite{1980A&A....87...63R} or single frequency images\cite{1986SoPh..107..159P} have demonstrated that Type III radio sources expand with time. This could
(for such single-frequency observations) be due to
either propagation effects or intrinsic variations in the structure of the Type~III burst\cite{1980IAUS...86..235P,1986SoPh..107..159P,1989A&A...210..417R}. Further, imaging observations with the Culgoora radioheliograph\cite{1972NPhS..238..115S,1985srph.book..289S} have also revealed an interesting enigma for the events at the limb: sources of fundamental emission are radially shifted outwards with respect to harmonic emission (and hence are apparently situated at different heights in the solar atmosphere), although the physics of the responsible coherent plasma emission mechanism requires that they are produced cospatially\cite{1980panp.book.....M}. This is particularly puzzling, since the refraction of radio waves shifts sources radially inwards and, since the fundamental component is refracted more than the second-harmonic component, the fundamental component should appear {\it lower}\cite{1985srph.book..289S}. Although a variety of possible resolutions of this paradox have been presented\cite{2002ApJ...575.1094W,2015ApJ...806...34W}, a possible resolution that is consistent with the observed increase in source size with time involves radio-wave propagation effects\cite{1972PASAu...2..100S,1999A&A...351.1165A}, which could\cite{1983PASAu...5..208R,1994kofu.symp..199B,2015MNRAS.447.3486I} shift the observed positions of fundamental radiation upward (radially outward). However, to date there have been no observations that permit the decoupling of propagation effects from intrinsic source variations, and hence there has been no quantitative assessment of the reasons behind this paradox.

Here we report imaging spectroscopy observations of fine frequency structures associated with a solar radio burst\cite{1972A&A....20...55D}. These observations with high spatial and temporal resolution demonstrate radio-wave propagation effects in the solar corona.

\section{Results}

\subsection{Overview of the observations}

 Radio burst on 2015~April~16 around 11:57 UT  was simultaneously observed by one of the largest decameter arrays, the LOw Frequency ARray (LOFAR)\cite{2013A&A...556A...2V} and by the URAN-2\cite{2016ExA....42...11K}
 (Ukrainian Radio interferometer of National Academy of Sciences).
 The latter provides corroborating observations at other frequencies, polarization information, and valuable cross-calibration for the LOFAR observations between 30 and 32~MHz.

\begin{figure}[pht]
\centering
\includegraphics[width=0.6\linewidth]{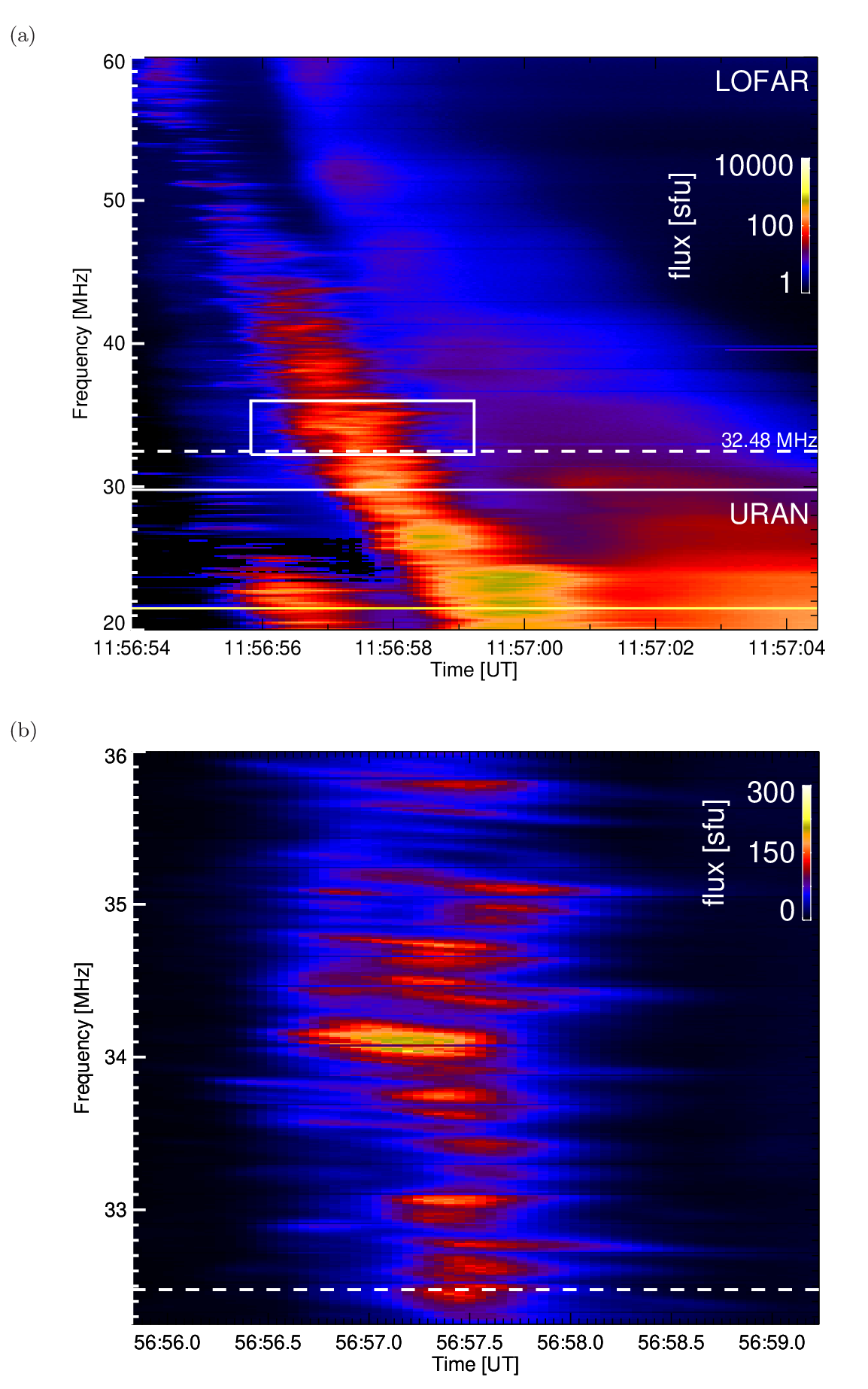}
\caption{\textbf{Sun-integrated dynamic spectrum of the solar radio burst.} (a) The Type III-IIIb solar radio burst observed on 2015~April~16 with both Low Frequency Arra (LOFAR)\cite{2013A&A...556A...2V}
and Ukrainian Radio interferometer of National Academy of Sciences (URAN-2)\cite{2016ExA....42...11K}. (b)  The expanded view of a 3-second interval shows finely-structured Type IIIb striae at frequencies between 32 and 36~MHz that have frequency widths of only $\sim$(0.1-0.3)~MHz.}
\label{fig:dynamic-spectrum}
\end{figure}

The dynamic spectrum (radio flux in the frequency-time plane; Figure~\ref{fig:dynamic-spectrum}) shows two main burst components, each characterized by a rapid decrease in frequency with time; the first burst passes through 20~MHz at $\simeq$11:57:00~UT and is followed a few seconds later by another burst which passes through 20~MHz at $\simeq$11:57:05~UT. The first burst is radiation at the fundamental plasma frequency, while the second burst is harmonic emission from the same electron beam forming a so-called type IIIb-type III pair\cite{1979SoPh...62..145A,2010AIPC.1206..445M}
(for example, at 11:57:00 the emission in the first burst is concentrated at frequencies around 20~MHz while the emission in the second burst is concentrated around 40~MHz.)

The Type III burst in this event is rather typical\cite{1985srph.book..289S,2010AIPC.1206..445M};
for example, the peak flux density between 32 and 40 MHz
is 100-200 solar flux units (sfu)
[1~sfu $=10^{-22}$~J~s$^{-1}$~m$^{-2}$~Hz$^{-1}$], and it has circular polarizations of $\sim$15\% and $<5$\% for the fundamental and harmonic components, respectively.   The rapid downward drift of frequency with time is a defining characteristic of solar Type III bursts\cite{1985srph.book..289S};
it results from the rapidly decreasing ambient density (and hence decreasing plasma frequency away from the Sun) as the emitting electron beam propagates upward through the decreasing density of the solar atmosphere.  Since the plasma frequency $f \propto n^{1/2}$, it follows that $(1/f) \, df/dt = (1/2n) \, dn/dt = (1/2) \, (d \ln n/dr) \, (dr/dt) = v/2L$, where $L =(d\ln n/dr)^{-1}$ is the density scale height and $v = dr/dt$ is the vertical component of the velocity of the exciting electron beam. Using the Newkirk\cite{1961ApJ...133..983N} density model of the solar corona as a typical model, the characteristic density scale height is $L \simeq 0.3R_\odot \simeq 2 \times 10^{10}$~cm at a level in the atmosphere corresponding to plasma frequencies around 32~MHz.  Therefore, the observed frequency drift rate $df/dt\simeq 7$~MHz~s$^{-1}$ at the $f=32$~MHz point in the fundamental frequency burst component corresponds to $v=(2L/f) (df/dt) \simeq 10^{10}$~cm~s$^{-1} \simeq c/3$, where $c$ is the speed of light. The speed $c/3$ is a typical speed for the $\sim 30$~keV electrons that excite Type~III bursts\cite{1985srph.book..289S,2015A&A...580A.137K}.

The expanded view in Figure~\ref{fig:dynamic-spectrum} shows that the fundamental component of the burst consists of multiple fine-structured striae; such fine structure is the characteristic signature of Type IIIb bursts\cite{1972A&A....20...55D} (the number of striae increases with decreasing frequency, so that below $\sim$30~MHz, the frequency structure of the burst looks quasi-continuous).
These fine frequency structures are believed to be due to small-scale density fluctuations\cite{1975SoPh...40..421T,2001A&A...375..629K,2010AIPC.1206..445M,2013SoPh..285..217R}
that modulate the resulting radio emission;
they have full width at half-maximum (FWHM) durations around 1~s at a given frequency (Figure~\ref{fig:dynamic-spectrum}). The presence of these fine striae in the fundamental component of the burst provides an estimate of the characteristic size of the emitting volume (intrinsic emission source size), which can then (see below) be compared to the source sizes obtained from direct imaging in order to evaluate the effects of radio-wave propagation on the observed source size.
Specifically, the individual striae (see the zoomed-in dynamic spectrum in Figure~\ref{fig:dynamic-spectrum}) have FWHM frequency widths $\Delta f \sim 0.3$~MHz.  Although the relationship between $\Delta f$ and the size of the radio emitting source is model dependent\cite{1982srs..work..182M}, and in particular depends on the angle between the direction of beam propagation and the direction of the density gradient,
an order-of-magnitude estimation based on the plasma emission mechanism suggests that a limited frequency range corresponds to a vertical extent $\Delta r \simeq 2L \, (\Delta f/f) \simeq 4 \times 10^8$~cm.
We note this would be the size of a density inhomogeneity
leading to an enhanced level of Langmuir waves,
while the electron beam generating the Langmuir waves is
extended over a much larger distance\cite{1975SoPh...40..421T,2001A&A...375..629K}.
Such a characteristic size of the fundamental
emitting source extends over an angle $\theta \simeq 0.1$~arcmin at the Sun and hence subtends a very small solid angle ($\Omega \simeq 10^{-2}$~arcmin$^2$) on the sky.
The harmonic emission is likely to form over a much larger region in physical space\cite{1989SoPh..120..369M}, a feature that is also evident from the dynamic spectra -- the fundamental component has clear striae, but the harmonic is rather smooth.

\subsection{Imaging}
LOFAR imaging observations were made using 24-core Low Band Antenna stations with tied-array beam forming\cite{2013A&A...556A...2V,2014A&A...568A..67M,2015A&A...580A..65M,2015A&A...579A..69O,2017A&A...606A.141R}, an observing mode that provides images with sub-second time resolution and unprecedented frequency resolution in order to resolve the individual striae in the Type IIIb burst. The array of 127 tied-array beams cover the sky out to $\sim$2$R_{\odot}$ with a mosaic beam spacing of $\sim0.1$~degrees. We note that tied array mosaic imaging is different from the traditional method of producing images from interferometric visibilities. The LOFAR core size of $D\simeq 3.5$~km provides an angular resolution $\lambda/D \simeq 9$~arcmin at 32~MHz (wavelength $\lambda =9.4$~m) and the ``dirty'' beam FWHM area $A_{\rm LOFAR}$ was $\simeq 110$~arcmin$^2$ at the time of the observation. The flux was calibrated against the Crab nebula both before and after the burst observations; in addition, the Sun-integrated flux was compared with URAN-2 data, which  showed agreement within a factor of $\sim$2. The temporal modulation of the URAN-2 flux also demonstrated excellent agreement with the observed fine-frequency structures, excluding  instrumental effects.

The imaging of the radio emission was performed with time resolution $\simeq$50~ms, during which radio waves propagate a distance of only $\sim$$1.5 \times 10^9$~cm~$ \simeq 0.3$~arcmin, allowing us to accurately track variations in both the location and the areal extent of the source on sub-second timescales. For each $12$~kHz-wide frequency channel we fitted an elliptical Gaussian to the LOFAR images. The ellipse centroid position (which is determined to an accuracy significantly better than the angular resolution of a single beam measurement\cite{1997PASP..109..166C}) and the FWHM area of each source were estimated for all frequency channels during the radio burst (Figure~\ref{fig:superimposed-images}).  Figure~\ref{fig:centroid-motion} shows the size and centroid positions (with uncertainties) of both fundamental (F) and harmonic (H) images for a typical stria near $32$~MHz;
the FWHM areas are $A^{\rm F}\sim 400$~arcmin$^2$ for the fundamental
and $A^{\rm H}\sim 600$~arcmin$^2$ for the harmonic.
The accuracy of determining the source position and area is variable depending on the emission flux (see Methods),
and near the burst peak they can be as high as $\pm 0.1$~arcmin for the position and $\pm 5$~arcmin$^2$ for the area
(see Figure \ref{fig:time-histories-flux-area}).
The areas and area uncertainties
are well above the LOFAR resolution limit
and  hence the radio sources are reliably resolved (see Methods).
The radio source sizes corrected for the LOFAR beam,
$A^{\{\rm F,H\}}-A_{\rm LOFAR}$, are both four orders of magnitude larger than the emission region size $\Omega\sim10^{-2}$ arcmin$^2$ determined above from considerations of the fine frequency width of individual burst striae.

Figure~\ref{fig:centroid-motion} shows the temporal evolution of the centroid location, and the areal extent, of the fundamental and harmonic sources (both observed at a frequency of 32.5~MHz, so that the H-radiation is produced in a region with a density one-fourth that of the region emitting the fundamental -- and a few seconds later, when the emission at the fundamental frequency has drifted downward to 16.25~MHz).  The centroid of the fundamental frequency radiation moves in a direction roughly parallel to the local solar radius (i.e., north-west in the plane-of-image; see full-disk image in Figure~\ref{fig:centroid-motion}), whereas the centroid of the source of harmonic radiation moves in a roughly transverse direction. The motion of F and H sources due to frequency drift between 38 and 32 MHz caused by electron transport is shown in Figure \ref{Fig:motion-all-sources}.
Figure~\ref{fig:time-histories-flux-area} shows the time evolution of the radial centroid positions and areas in the X-Y plane, for both F and H components (Figure~\ref{fig:centroid-motion}).
The areal expansion of both F and H components is most pronounced during the decay of the burst. This is consistent with various wave scattering models\cite{1971A&A....10..362S,1983PASAu...5..208R,1999A&A...351.1165A}, although these models predict different motions and growth rates of the source, depending on the assumed emission and scattering anisotropies\cite{1977A&A....61..777B}.  Therefore, we focus on times during the decay and estimate the radial velocity in the X-Y plane and areal expansion rate by fitting linear expressions $r(x,y) = r_0 + (dr/dt) (t-t_0), A = A_0 + (dA/dt) (t-t_0)$ during the time intervals shown by the shaded regions in Figure~\ref{fig:time-histories-flux-area}. The centroid of the F-emission moves radially outward at an average speed $dr/dt \simeq 1.8$~arcmin~s$^{-1} \simeq c/4$,
while its area $A^{\rm F}$ grows from $\sim$420~arcmin$^{2}$ to $\sim$530~arcmin$^{2}$ within $\sim 0.6$~sec,
an average areal expansion rate $dA^{\rm F}/dt \simeq 180$~arcmin$^2$~s$^{-1}$. On the other hand, the centroid of the harmonic component shows negligible radial motion, while its area $A^{\rm H}$ grows from $\sim$600~arcmin$^{2}$ to $\sim$760~arcmin$^{2}$ over $\sim3$~s. The average areal expansion rate is $dA^{\rm H}/dt \simeq 50$~arcmin$^2$~s$^{-1}$,
about one-fourth the areal expansion rate for the fundamental component.

We repeated this analysis for 48 well-observed striae in frequency channels between 32 and 38~MHz (Figure \ref{fig:average-values}). We excluded frequencies below 32~MHz where the striae start to overlap and images above 38~MHz due to low signal-to-noise ratio. Figure~\ref{fig:average-values} shows that for all well-resolved striae the rate of areal expansion of the fundamental source is $\sim$(2-4) times greater than the expansion rate for the harmonic source.

Individual striae start at different times within the Type IIIb burst (Figure~\ref{fig:dynamic-spectrum}). Each stria also initially appears at a different location on the solar disk within a broad envelope of the Type IIIb burst (see Methods). However, nearly all stria sources move radially while the harmonic component at the same frequency behaves in a completely different manner. This allows us to exclude refraction effects in the Earth's ionosphere as an explanation for the observed motion of the fundamental stria component.  Further, the very similar expansion rates inferred from observations in 235 different frequency channels, at 48 different striae spread over 2~s (larger than $\sim$1~s duration of a stria) allow us to infer with a high degree of confidence that {observed regions of fundamental radiation expand faster than regions of harmonic radiation}.  As we argue next, this result is not supported by any reasonable variation in the intrinsic source sizes in existing models\cite{2008A&ARv..16....1P}, but is consistent with propagation-scattering effects.

An intrinsic variation of source size with time at a given frequency requires that the emitting source grows in time as larger and larger iso-density surfaces start to emit\cite{1989A&A...210..417R}. However, in order to produce the observed striae, which are very narrow in frequency and large in imaged sources, such a model would require two essential features.
Firstly, the emitting region is distributed over a thin but large (and changing) volume, all at the same plasma frequency and thus density; any density inhomogeneities would have to be always parallel to iso-frequency surfaces, and secondly, the positions of the stria, which originate at different locations, have nearly identical centroid motions. Moreover, in such a scenario, the expansion rate is related to the structure of the iso-density surface and it is challenging to explain why the expansion has a similar rate at all frequencies, or equivalently why all the iso-density surfaces, which are spread over a height range $\sim 0.2R_{\odot}$, expand at nearly identical rates. Finally, such a model does not explain why the centroids of the fundamental and harmonic components behave differently. Therefore, we reach the rather inescapable conclusions that the emitting sources most probably have sizes comparable to the inhomogeneity scale; they are randomly located within the corona and are responsible for the individual striae; and that the observed extent of the radio burst is primarily determined not by the size of the emitting region but rather by wave propagation effects in the surrounding atmosphere.

\subsection{Radio wave propagation}
A simple model\cite{1971A&A....10..362S,1999A&A...351.1165A} for multiple scattering off density inhomogeneities  does indeed produce a (linear) increase in the area of a source (in solid angle units) with time (see Equation 64 from\cite{1999A&A...351.1165A}): $\frac{d<\theta ^2>}{dt} \propto \frac{f_{\rm pe}^4}{f^4} \, \frac{1}{\mu^4} \, \frac{\langle\delta n^2\rangle}{n^2\ell}$, where ${\langle\delta n^2\rangle}$ is the rms level of density fluctuations, $\ell$ is the density inhomogeneity scale\cite{1999A&A...351.1165A,2016JGRA..12111605S}, and $\mu$ is the refractive index.  For a fully ionized plasma, $\mu^2(f)=1-f_{\rm pe}^2/f^2$ is frequency dependent, so the rate of areal increase for radiation near the plasma frequency will always be larger than that for radiation at the harmonic frequency, qualitatively consistent with the observations in Figure~\ref{fig:average-values}.  As the waves propagate away from the source into regions of lower density (and so lower plasma frequency), the (local) plasma frequency $f_{\rm pe}$ becomes progressively smaller than the wave frequency $f$, the refractive index approaches unity for both fundamental and harmonic radiation, and the expansion rate $\propto f_{\rm pe}^4/\mu^4$ is greatly reduced.
The location of the radio source is $\sim 7$~arcmin from the solar disk center, so the short temporal extent (average FWHM $\Delta t \simeq 1.1$~s) of the striae constrains the radio wave broadening along the line of sight direction to be less than $c \Delta t \simeq 3 \times 10^{10}$~cm~$\simeq 8$~arcmin.
Since the perpendicular to line of sight size ($\sim$20~arcmin on the plane of the sky) is larger than the line of sight $8$~arcmin size,
the combined effect of scattering and emission directivity\cite{1980panp.book.....M}
is likely to be anisotropic, with the dominant effect being perpendicular to the line-of-sight direction. While both the fundamental and harmonic regions expand, the fundamental source is also radially moving, and this suggests a rather small intrinsic source size for the fundamental emission and a somewhat larger intrinsic source size for the harmonic emission. Indeed, the harmonic emission source is $\sim$1.4 times larger than the fundamental (Figure \ref{fig:time-histories-flux-area}).

\section{Discussion}

We have reported imaging spectroscopy observations of fine frequency structures in solar Type~III radio bursts. The high frequency-time resolution of these observations have allowed us to image the radio-waves as they emerge from the solar atmosphere.

\begin{figure}[htbp]
\centering
\includegraphics[width=0.5\linewidth]{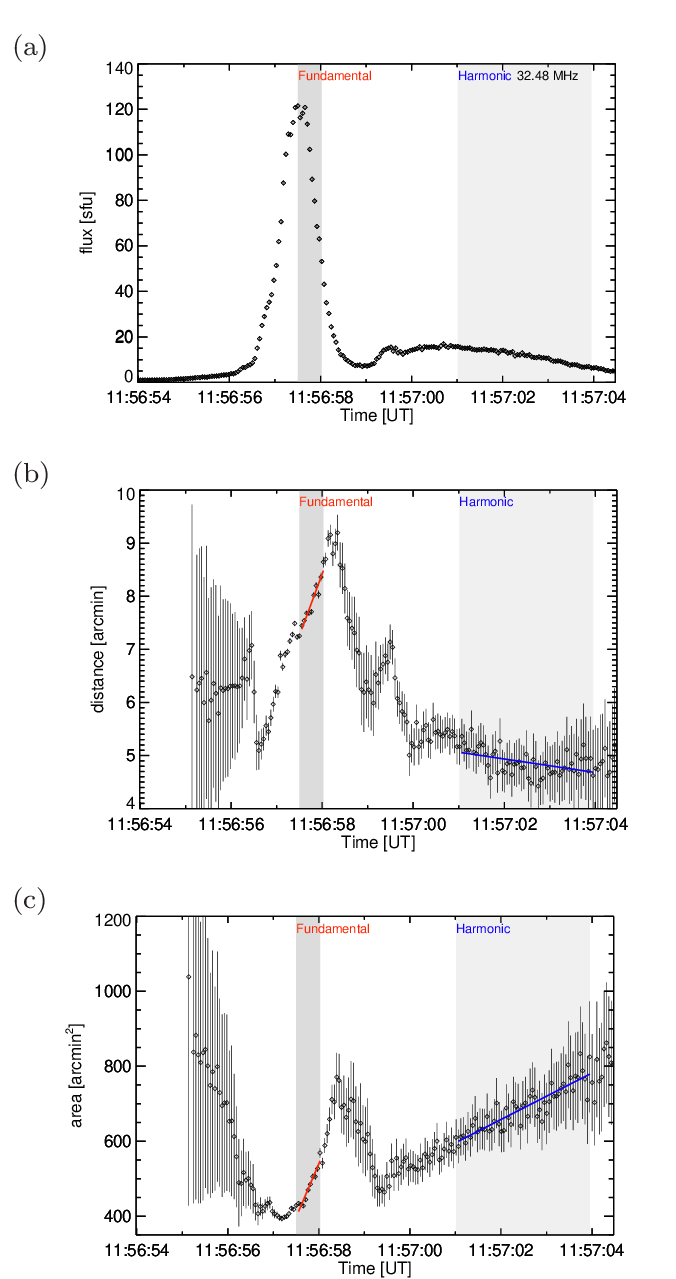}
\caption{\textbf{Time variations of flux, radial distance from the Sun center, and areal extent, for the selected stria in the 32.5~MHz frequency channel.} (a) time histories of the F- and H- components of the radio flux density in solar flux units (sfu) (b) Radial distances of the F and H sources versus time; (c) Areas of the F- and H-source areas versus time. Linear fits (red and blue lines, for the F and H sources, respectively) to the radial positions $r=r_0 + (dr/dt) \, (t-t_0)$ and areas $A = A_0 + (dA/dt) \, (t-t_0)$ were applied in the time ranges shown by the dark and light grey patches, respectively. The error bars represent one standard deviation of uncertainty. The uncertainties of the source size and position were determined by the 2D Gaussian fit (see Methods).}
\label{fig:time-histories-flux-area}
\end{figure}

\begin{figure}
\centering
\includegraphics[width=0.5\linewidth]{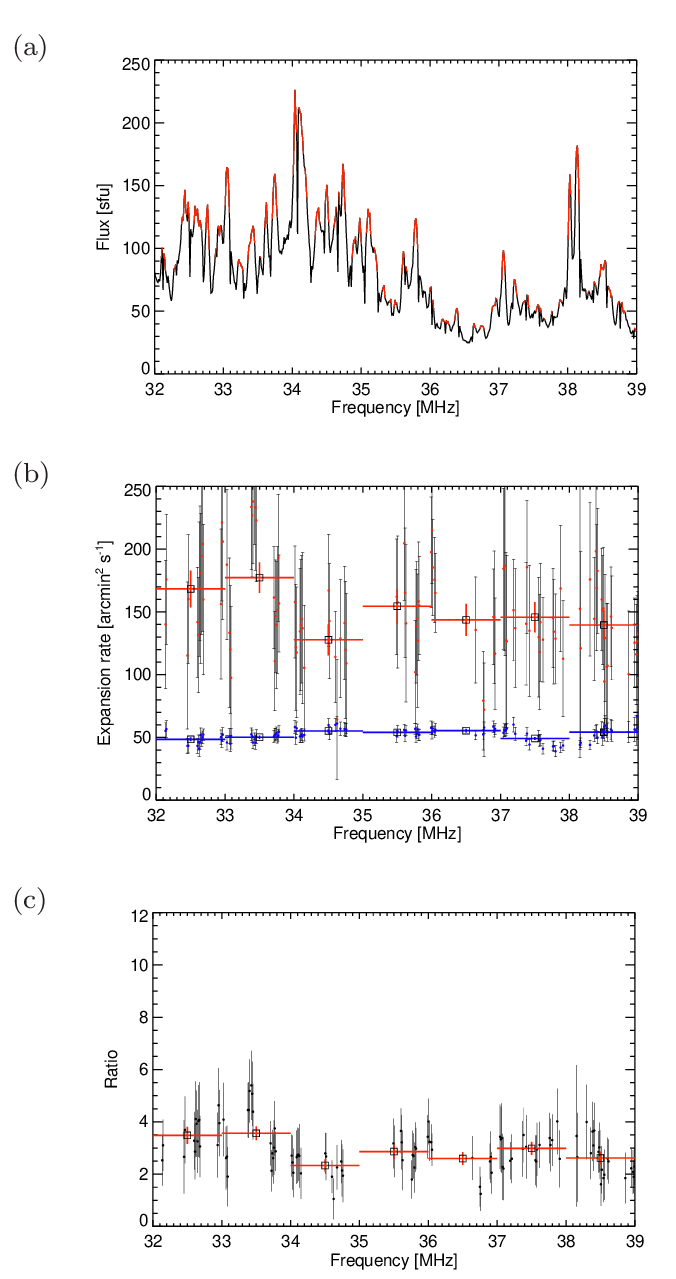}
\caption{\textbf{Statistical properties of the source areal expansion factors.} (a) Flux along the spine of the Type IIIb burst as a function of frequency. Peaks colored in red indicate selected fine temporal stria. (b) Expansion rate $dA/dt$ for all frequency channels with well-observed striae (those indicated by red in the top panel); the red and blue lines show the 1-MHz average values for fundamental and harmonic radiation, respectively. (c) ratio of the expansion rates $(dA^{\rm F}/dt)/(dA^{\rm H}/dt)$ as a function of frequency, averaged over the 1~MHz frequency bins. The error bars represent one standard deviation of uncertainty. }
\label{fig:average-values}
\end{figure}

The observed sources have linear extents, corrected for the finite size of the LOFAR beam, of $\sim \sqrt{(A^{\rm F}-A_{\rm LOFAR})}\simeq (17-22)$~arcmin near the peak of the fundamental component (Figure~\ref{fig:time-histories-flux-area}). The source sizes are very similar to the average source sizes $\simeq$20~arcmin inferred for Type~III bursts at 43~MHz\cite{1985srph.book..289S}
and are much larger than the intrinsic emission source sizes $\sim$$0.1$~arcmin deduced from the appearance of fine temporal substructures within the burst. The simulations of LOFAR response (see Methods) show the small source required by plasma emission cannot explain the LOFAR observations. We have also found that the areal extent of the fundamental component grows more rapidly than the harmonic component, consistent with a model involving scattering off density inhomogeneities.  The measured expansion rates (Figure~\ref{fig:average-values}),
as well as the sub-second fundamental source motion, provide valuable information on the fibreous structure of the corona\cite{1983PASAu...5..208R} and on the (currently poorly known) characteristics of the density turbulence spectrum\cite{1971A&A....10..362S,1999A&A...351.1165A}.

Because the intrinsic source size is so much smaller than the apparent source sizes, the brightness temperature of the source must be similarly ($A^{\rm F}/\Omega \sim 10^4$) larger than that obtained using the apparent source areas. A flux of 100 solar flux units at the fundamental frequency corresponds to a brightness temperature $\sim 10^{14}$~K, which is larger than what is typically assumed\cite{2008A&ARv..16....1P} and interestingly close to the maximum brightness temperatures observed for type III solar radio bursts\cite{1989SoPh..120..369M}.

\begin{figure}
\centering
\includegraphics[width=0.7\linewidth]{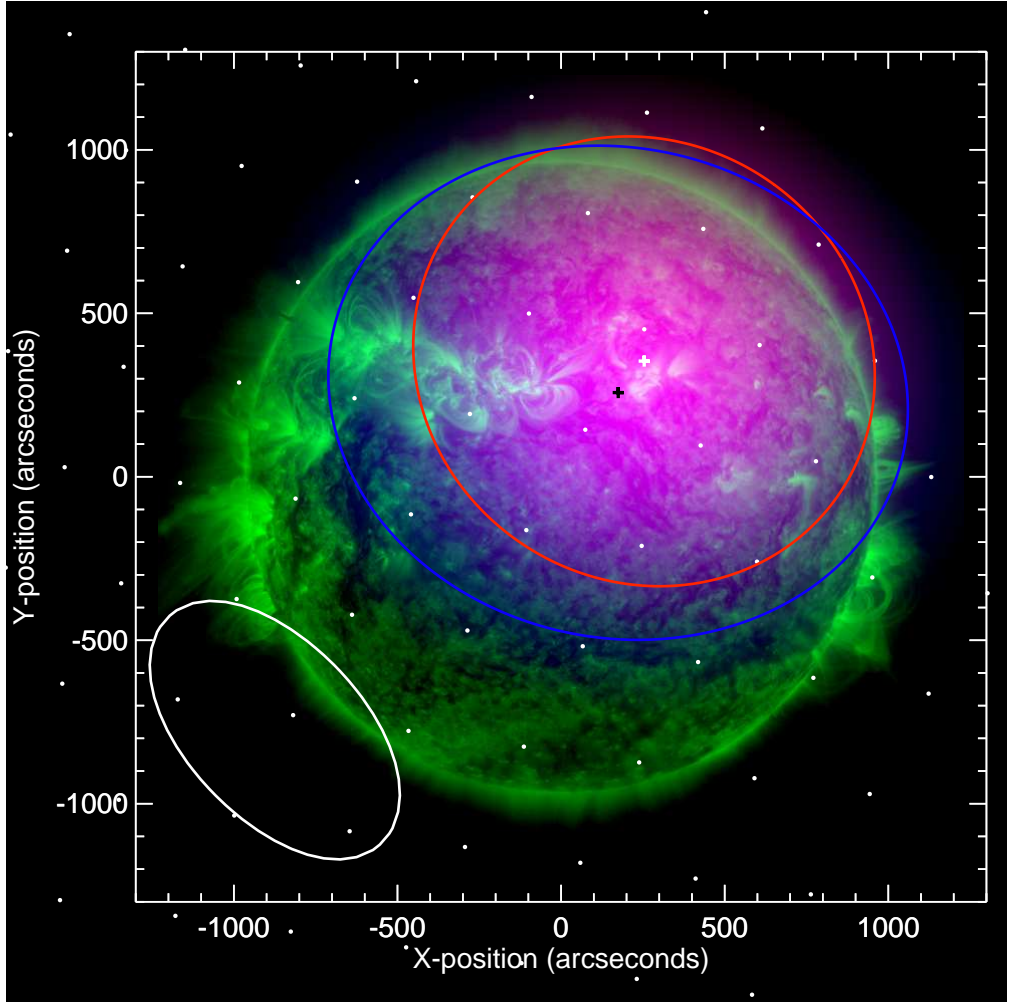}
\caption{\textbf{Radio images of the fine structure components of the burst.} Superimposed images of the Extreme Ultra-Violet (EUV)
and radio emission at the selected 32.5~MHz frequency
(Figure~\ref{fig:dynamic-spectrum}). {\it Green}: Observations from the Solar Dynamics Observatory/Atmospheric Imaging Assembly\cite{2012SoPh..275...17L} {171~\AA}
at 2015 April 16 11:57 UT;
{\it Red}: radio fundamental plasma frequency (F) component at 11:56:57.5 UT; {\it Blue}: second harmonic (H) radio component at 11:57:01 UT. The centroid positions of the F- and H-components are marked with white and black crosses, respectively. The full width at half maximum (FWHM) ellipses are made using two-dimensional Gaussian fits to the data. The white dots show the phased array beam locations and the oval shows the half-maximum synthesised Low Frequency Array (LOFAR) beam. See also Supplementary Movie 1.}
\label{fig:superimposed-images}
\end{figure}

\begin{figure}
\centering
\includegraphics[width=0.76\linewidth]{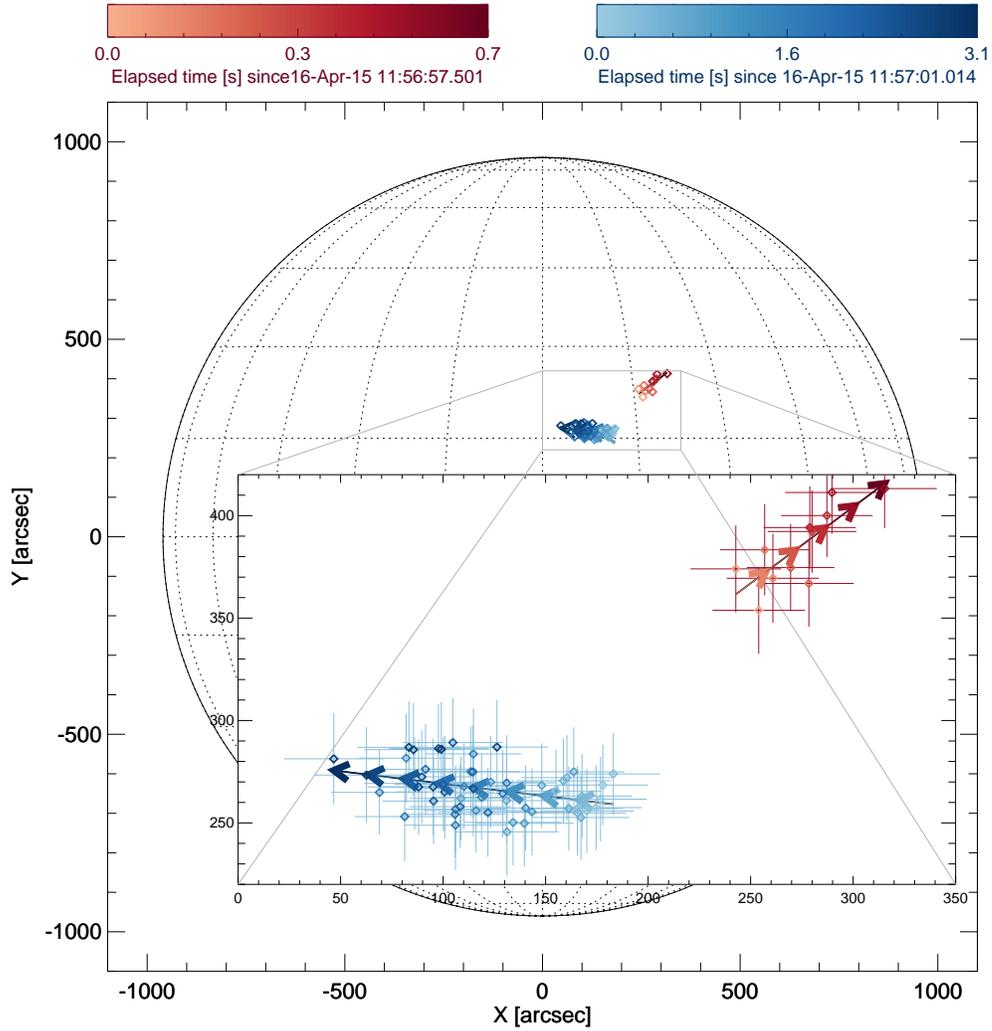}
\caption{\textbf{Centroid locations of the fundamental (F; red) and harmonic (H; blue) sources} for 32.5~MHz as a function of time, determined using a two-dimensional Gaussian fit to each observed source. Darker colors correspond to later times, as shown in the color scale in the insert. Straight-line fits to the positions of each centroid are shown by the arrows. The time elapsed is measured as time after the flux peak;
11:56:57.6~UT for fundamental and 11:57:01~UT for harmonic
(Figure~\ref{fig:time-histories-flux-area}). The full solar disk shows clearly that the F source is displaced radially outwards. The error bars represent one standard deviation of uncertainty. The uncertainties of the source position were determined by the 2D Gaussian fit (see Methods).}
\label{fig:centroid-motion}
\end{figure}

These results also resolve a long-standing problem in solar radio astronomy -- why the fundamental and harmonic sources, which result from the same physical process, do not generally appear to coincide spatially (Figures~\ref{fig:superimposed-images} and~\ref{fig:centroid-motion}). Because propagation effects result in a large increase in source size with time, the apparent source locations of the F and H sources are controlled primarily by these propagation-scattering effects, rather than by properties of the underlying emitting sources. We encourage further tests of this conclusion through analysis of radio data in different frequency bands.

The differential rates of propagation of F- and H-radiation can easily lead, within a modest time, to a systematic displacement of the apparent source centroids, which Figure~\ref{fig:centroid-motion} shows to be in the radial direction.

These observations also allow new testing and improvement
of radio wave propagation models\cite{1983PASAu...5..208R,1999A&A...351.1165A,2008ApJ...676.1338T} in turbulent coronal plasma, tests that have hitherto not
been available for the solar corona\cite{2016JGRA..12111605S}. In particular, since the areal expansion coefficient is proportional to the size of the density fluctuations, this provides a diagnostic of the latter quantity and opens up previously unavailable opportunities for further study of various manifestations of solar activity, such as flares, coronal mass ejections and formation of the solar wind.

\begin{figure}
\centering
\includegraphics[width=0.8\linewidth]{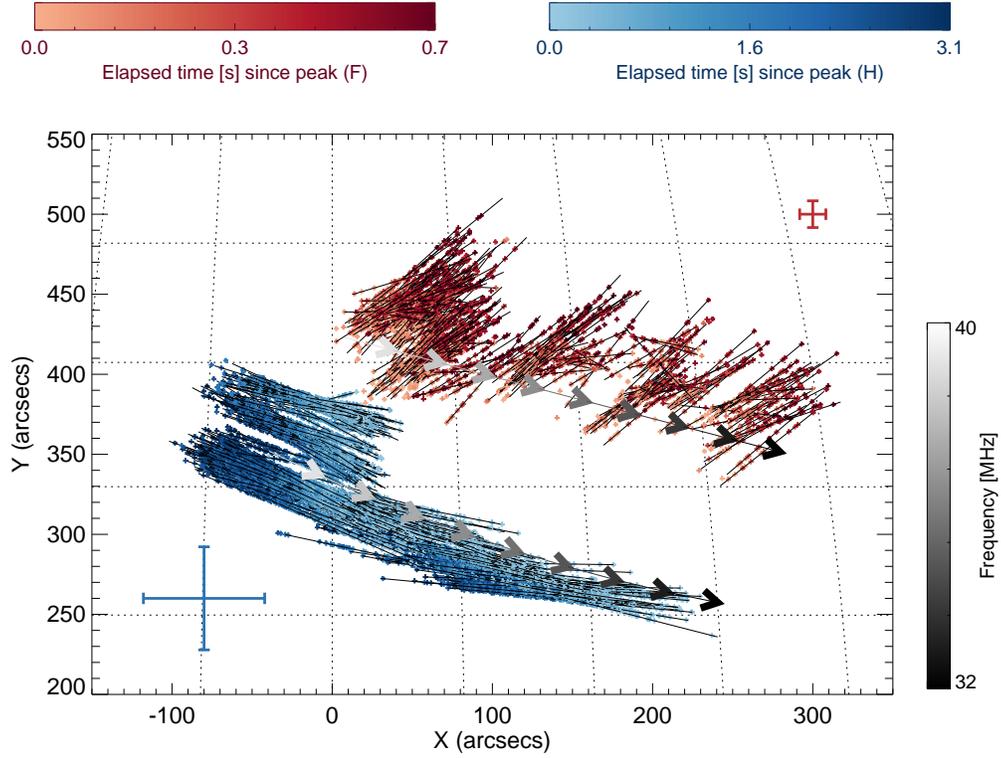}
\caption{\textbf{Motion of the sources}. Grey arrows show the projected motion of the burst component in frequency as the burst drifts in frequency given by the color bar. Centroid positions as a function of time for fundamental (red) and harmonic (blue) components with time after the peak at each frequency (see Figure \ref{fig:time-histories-flux-area}).
The error bars represent one standard deviation of uncertainty. The uncertainties of the position determined by the 2D Gaussian fit (see Methods) are given by the red and blue crosses. }
\label{Fig:motion-all-sources}
\end{figure}

\begin{figure}
\centering
\includegraphics[width=0.5\linewidth]{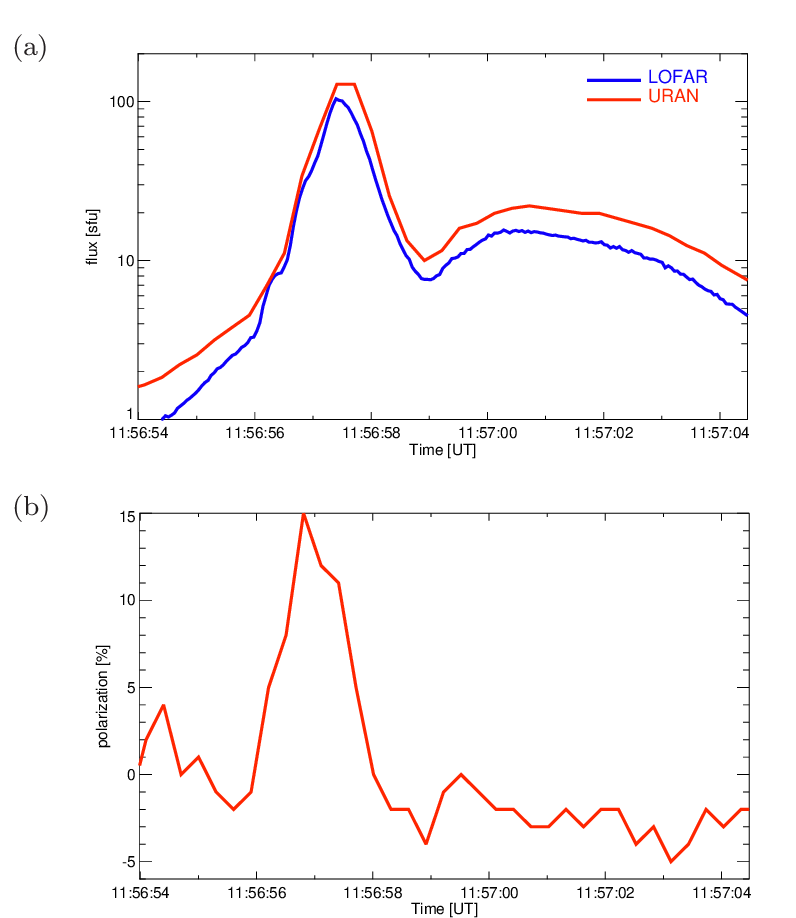}
\caption{\textbf{Total flux and polarization data.}
(a) Time profile of the Sun-integrated radio flux at 32~MHz from the Low Frequency Array (LOFAR) and
Ukrainian Radio interferometer of National Academy of Sciences (URAN-2). (b) Degree of circular polarization of the radio emission at the same frequency.}
\label{Fig:cross-calibration}
\end{figure}

\begin{figure}
\centering
\includegraphics[width=0.9\linewidth]{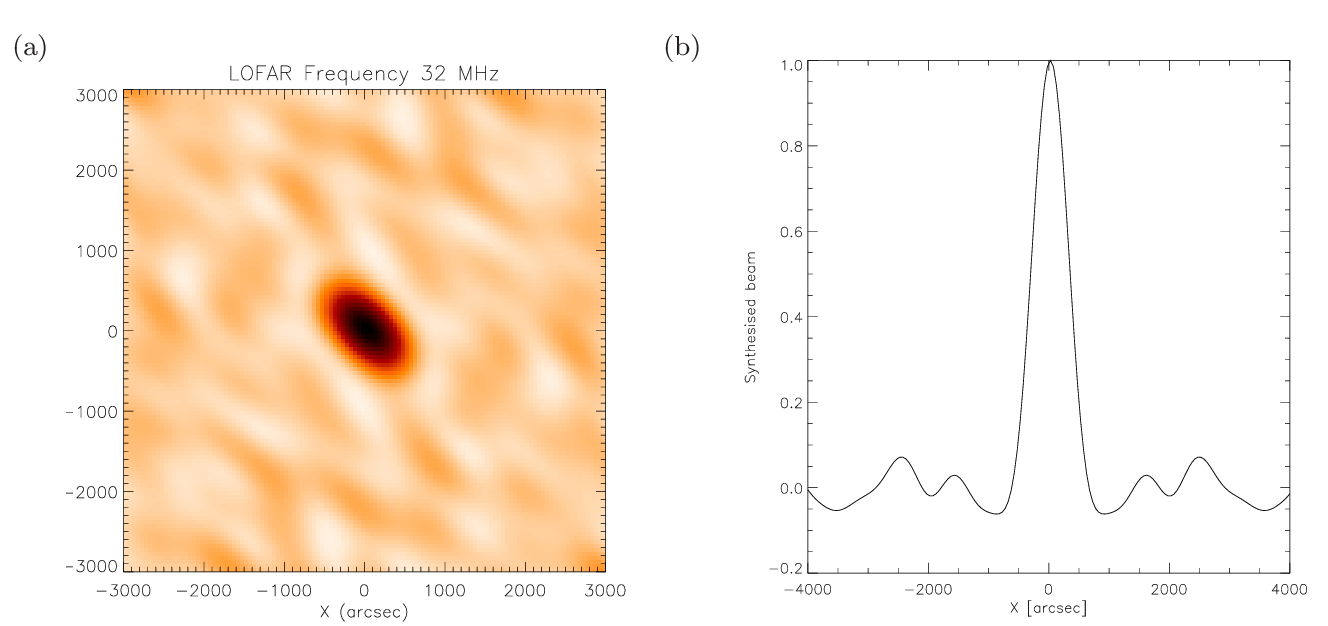}
\caption{\textbf{The synthesized tied-array beam from 24 Low Frequency Array (LOFAR) stations at 32 MHz }. (a) 2D synthesised beam centered at the Sun centre. (b) The slice of the beam at $y=0$.
Full Width at Half Maximum (FWHM) area of the synthesized is around $110$~arcmin$^2$.}
\label{fig:psf}
\end{figure}

\begin{figure}
\centering
\includegraphics[width=0.8\linewidth]{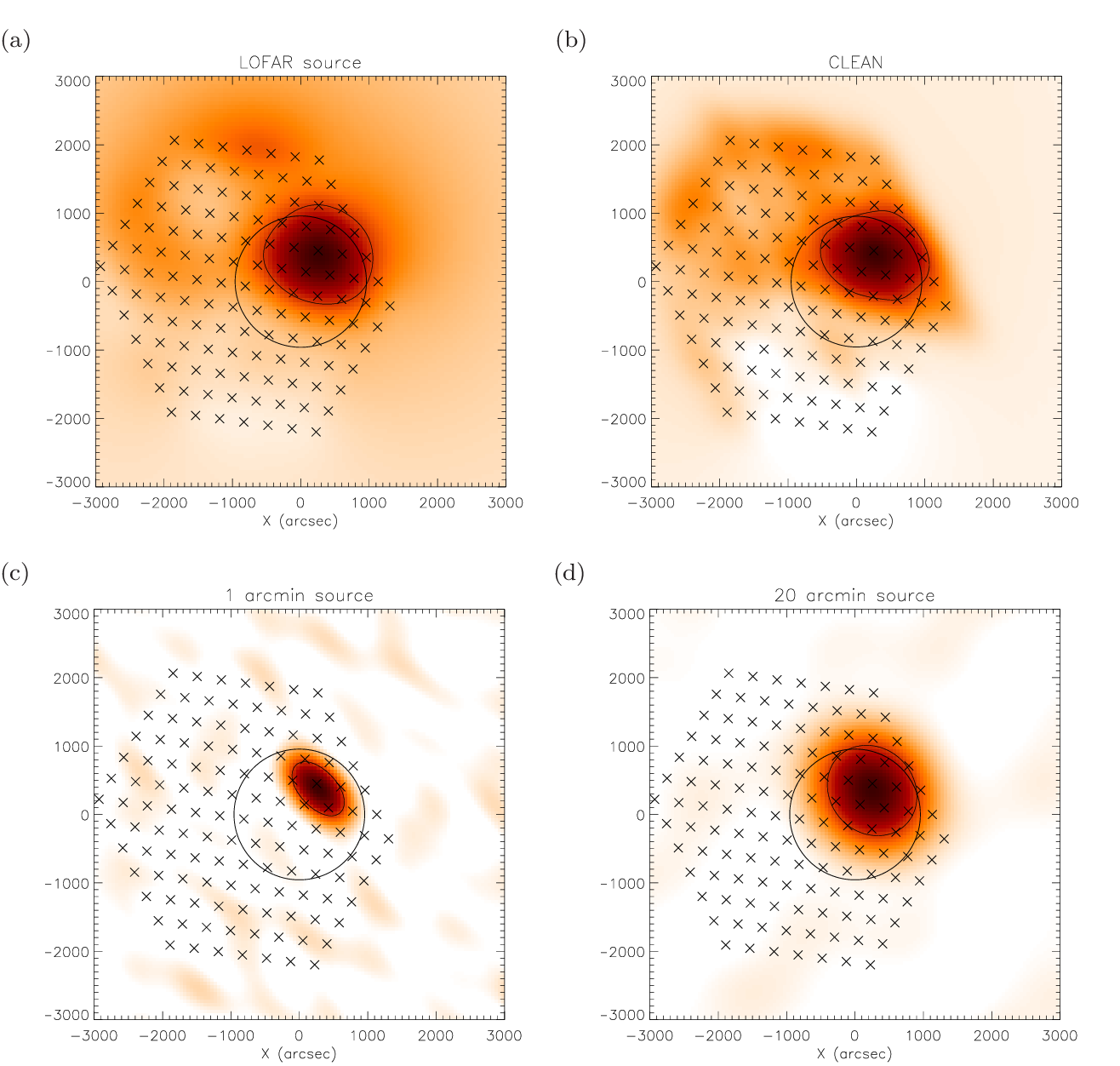}
\caption{\textbf{Image cleaning and simulation of Low Frequency Array (LOFAR) response}.
(a) The `Dirty' image resulting from linear interpolation between the beam locations shown as crosses; (b) CLEANed image for the same dataset as top right; The bottom row demonstrates the simulated LOFAR images from a source located $x=200''$, $y=400''$ and convolved with the LOFAR beam; (c) compact  1~arcmin Full Width at Half-Maximum (FWHM) circular source; (d) extended 20 arcmin FWHM circular source. The thin solid line in all images shows 50\% intensity level. Bold circle shows the Sun and the beam positions are given by crosses. All images are normalised to 1 for easy comparison.}
\label{fig:model_clean}
\end{figure}
\section{Methods}
\subsection{Polarization}
The Sun-integrated flux densities observed by LOFAR and URAN-2  described above has been compared in Figure \ref{Fig:cross-calibration}. Polarization measured by URAN-2 is given for the same frequency.

\subsection{Radio source position and size}
The imaging has been performed for 48 individual stria in 235 frequency channels and is shown in Figure~\ref{Fig:motion-all-sources}. The LOFAR observations were made using 24-core Low Band Antenna stations using tied-array beam forming images with sub-second time resolution. The array of 127-tied array beams, which are the coherent sum of all the station beams, covered the Sun with $\sim6$~arcmin spacing (the data are publicly available via http://lofar.target.rug.nl).
Each LOFAR beam measures the flux $F_i$ for a given position on the sky given by $x^b_i$, $y^b_i$ in heliocentric coordinates.
Figure \ref{fig:superimposed-images} shows the locations of the beams on the solar disk. The measured flux $F_i$ recorded by LOFAR is a convolution of a LOFAR beam point spread function (Figure \ref{fig:psf}) and the true source. The positions and the flux values for each beam allows us to reconstruct the images. Iteratively finding the highest value in the array of $F_i$, 1\% of the maximum convolved with a point spread
function (Figure \ref{fig:psf}) has been subtracted from all $F_i$
until the highest value is smaller than 5\% of the initial maximum.
The resulting cleaned image\cite{1974A&AS...15..417H,1995ASPC...75..318K} is shown
in Figure \ref{fig:model_clean}. Figure \ref{fig:model_clean} also compares the result of simulated LOFAR images.

In principle, one can apply the cleaning procedure to thousands of images, and then deduce the measured source parameters.
However, because we are interested in the characteristic size and source position only, such an extensive calculation effort seems unjustified.
Instead,  we determine the areal extent and subtract the LOFAR dirty
beam area. The side-lobes do not exceed 10\%, so the images are weakly affected by cleaning at half maximum level (the top right (cleaned) image looks similar to the top left (dirty) image at 50\% level in Figure \ref{fig:model_clean}). The simulations in Figure \ref{fig:model_clean} also demonstrates that small $\simeq 1$ arcmin sources are inconsistent with the data.

To determine the size and the position of the source,
we fit an elliptical Gaussian $S(x,y)=S_0
\exp (-x'^2/2 \sigma_x ^2-y'^2/2 \sigma_y ^2)$,
where $x'=(x-x_{\rm s})\cos(T)-(y-y_{\rm s})\sin(T)$ and $y'=(x-x_{\rm s})\sin(T)+(y-y_{\rm s})\cos(T)$, where $T$ is the rotation
from the X-axis, in the clockwise direction.
Minimizing $\chi ^2$
\[
\chi ^2 =\sum _{i=1}^{N}\frac{(F_i-S(x^b_i, y^b_i; S_0, x_{\rm s},y_{\rm s},\sigma_x,\sigma_y,T))^2}{\delta {F}^2}\,,
\]
we find the Gaussian parameters for each moment of time and frequency. The inferred parameters are as follows: $S_0$ is the peak amplitude, $x_{\rm s},y_{\rm s}$ are the coordinates of the central source position, $\sigma_x,\sigma_y$ are the rms lengths,
$T$ is the rotation of the ellipse from the $x$ axis.
The background flux level before the burst
 for each frequency and beam was taken as the uncertainties on the flux $\delta F$ (typically around 1 sfu). The resulting fitted Gaussian gives the source size (convolved with the LOFAR beam).
The half-maximum area of the fit gives the areal extent of the source. Since we are interested in the position and the areal extent
of the source we provide the expressions for the errors.
The errors $\delta {x_{\rm s}}$, $\delta {y_{\rm s}}$
for the source position ($x_{\rm s}$, $y_{\rm s}$) can be written\cite{1997PASP..109..166C}
\[
\delta {x_{\rm s}}\approx \sqrt{\frac{2}{\pi}}\frac{\sigma _x}{\sigma_y}
\frac{\delta F}{S_0}h,\;\;\;
\delta {y_{\rm s}}\approx \sqrt{\frac{2}{\pi}}\frac{\sigma _y}{\sigma_x}
\frac{\delta F}{S_0}h,
\]
where $h$ is the angular resolution. For weakly elliptical source $\sigma _x \sim \sigma _y$ as presented in
Figure~\ref{fig:superimposed-images},
the uncertainty on the source position gives $\delta y_{\rm s} \simeq \delta {x_{\rm s}}\sim 0.1$~arcmin for the fundamental ($\sim 120$ sfu) and
$\delta {x_{\rm s}}\simeq \delta {y_{\rm s}}\sim 0.6$~arcmin
for the harmonic since the harmonic ($\sim 17$ sfu)
has 5-6 times lower flux near the peak of each component
(Figure~\ref{fig:time-histories-flux-area}).

Similarly, the error on the source half-maximum area $A$ can be estimated
 \[
 \frac{\delta A}A \approx 2 \frac{\delta F}{S_0}\frac{h}{\sqrt{A}}
 \]
 For example, the error on the area becomes $\delta A/A\sim 5$~\% for harmonic near the peak shown in Figure \ref{fig:time-histories-flux-area}. The above expressions show that
 the accurate determination of the source positions and the areas
becomes available due to high signal to noise ratios $F_i/\delta F$.
We further note that the results do not
imply that the source has a Gaussian shape, but show that the
multi-beaming measurements can provide estimates
of the position better that the resolution of the instrument
and the area measurements better that 'dirty' beam
half-maximum area.

\noindent \textbf{Data availability:} The datasets generated during and/or analysed during the current study are available in the LOFAR Long Term Archive, http://lofar.target.rug.nl/
and {https://sdo.gsfc.nasa.gov/data/}
or available from the authors upon request.

\section*{Acknowledgements}
 E.P.K., N.L.S.J., N.H.B. were supported by a STFC consolidated grant ST/L000741/1. A.G.E. was supported by grant NNX10AT78G from  NASA’s Goddard Space Flight Center. A.A.K. was supported in part by the RFBR grant 15-02-03717. B.A. was supported
     by a STFC studentship grant. The work has benefited from
     a  Marie  Curie  International  Research Staff Exchange Scheme 'Radiosun' (PEOPLE-2011-IRSES-295272) and
     an international team grant (http://www.issibern.ch/teams/lofar/)
     from ISSI Bern, Switzerland. This paper is based (in part) on data obtained with the International LOFAR\cite{2013A&A...556A...2V} Telescope (ILT). LOFAR is the Low Frequency Array designed and constructed by ASTRON. It has facilities in several countries, that are owned by various parties (each with their own funding sources), and that are collectively operated by the ILT foundation under a joint scientific policy.

\bibliographystyle{h-physrev3}
\bibliography{all_issi_references3}

\begin{thebibliography}{10}

\bibitem{2011SSRv..159..107H}
G.~D. {Holman} {\em et~al.},
\newblock \ssr {\bf 159}, 107 (2011), 1109.6496.

\bibitem{2008A&ARv..16....1P}
M.~{Pick} and N.~{Vilmer},
\newblock \aapr {\bf 16}, 1 (2008).

\bibitem{1980panp.book.....M}
D.~B. {Melrose},
\newblock {\em Plasma astrophysics. Nonthermal processes in diffuse magnetized
  plasmas.} (New York: Gordon and Breach, 1980, 1980).

\bibitem{1975SoPh...40..421T}
T.~{Takakura} and S.~{Yousef},
\newblock \solphys {\bf 40}, 421 (1975).

\bibitem{1982srs..work..182M}
D.~B. {Melrose},
\newblock {Fine structures in Decametric noise storms : possible mechanisms},
\newblock in {\em Solar Radio Storms, CESRA Workshop \#4}, edited by A.~O.
  {Benz} and P.~{Zlobec}, p. 182, 1982.

\bibitem{1971A&A....10..362S}
J.~L. {Steinberg}, M.~{Aubier-Giraud}, Y.~{Leblanc}, and A.~{Boischot},
\newblock \aap {\bf 10}, 362 (1971).

\bibitem{1965BAN....18..111F}
A.~D. {Fokker},
\newblock \bain {\bf 18}, 111 (1965).

\bibitem{1974SoPh...35..153R}
A.~C. {Riddle},
\newblock \solphys {\bf 35}, 153 (1974).

\bibitem{1972PASAu...2..100S}
R.~T. {Stewart},
\newblock Proceedings of the Astronomical Society of Australia {\bf 2}, 100
  (1972).

\bibitem{1983PASAu...5..208R}
R.~D. {Robinson},
\newblock Proceedings of the Astronomical Society of Australia {\bf 5}, 208
  (1983).

\bibitem{1999A&A...351.1165A}
K.~{Arzner} and A.~{Magun},
\newblock \aap {\bf 351}, 1165 (1999).

\bibitem{1994ApJ...426..774B}
T.~S. {Bastian},
\newblock \apj {\bf 426}, 774 (1994).

\bibitem{2008ApJ...676.1338T}
G.~{Thejappa} and R.~J. {MacDowall},
\newblock \apj {\bf 676}, 1338 (2008).

\bibitem{1980A&A....87...63R}
A.~{Raoult} and M.~{Pick},
\newblock \aap {\bf 87}, 63 (1980).

\bibitem{1986SoPh..107..159P}
M.~{Pick} and S.~C. {Ji},
\newblock \solphys {\bf 107}, 159 (1986).

\bibitem{1980IAUS...86..235P}
M.~{Pick}, A.~{Raoult}, and N.~{Vilmer},
\newblock {Observations of solar type III radio bursts with the Nancay
  radioheliograph},
\newblock in {\em Radio Physics of the Sun}, edited by M.~R. {Kundu} and T.~E.
  {Gergely}, , IAU Symposium Vol.~86, pp. 235--240, 1980.

\bibitem{1989A&A...210..417R}
E.~C. {Roelof} and M.~{Pick},
\newblock \aap {\bf 210}, 417 (1989).

\bibitem{1972NPhS..238..115S}
K.~V. {Sheridan}, N.~R. {Labrum}, and W.~J. {Payten},
\newblock Nature Physical Science {\bf 238}, 115 (1972).

\bibitem{1985srph.book..289S}
S.~{Suzuki} and G.~A. {Dulk},
\newblock {Bursts of Type III and Type V},
\newblock in {\em Solar Radiophysics: Studies of Emission from the Sun at Metre
  Wavelengths}, edited by D.~J. {McLean} and N.~R. {Labrum}, pp. 289--332,
  Cambridge University Press, 1985.

\bibitem{2002ApJ...575.1094W}
C.~S. {Wu}, C.~B. {Wang}, P.~H. {Yoon}, H.~N. {Zheng}, and S.~{Wang},
\newblock \apj {\bf 575}, 1094 (2002).

\bibitem{2015ApJ...806...34W}
C.~B. {Wang},
\newblock \apj {\bf 806}, 34 (2015), 1504.01126.

\bibitem{1994kofu.symp..199B}
T.~S. {Bastian}, N.~{Nitta}, A.~L. {Kiplinger}, and G.~A. {Dulk},
\newblock {Energy Transport During a Solar Flare: VLA Observations of the M1.9
  Flare of 20 Aug 1992},
\newblock in {\em Proceedings of Kofu Symposium}, pp. 199--202, 1994.

\bibitem{2015MNRAS.447.3486I}
M.~{Ingale}, P.~{Subramanian}, and I.~{Cairns},
\newblock \mnras {\bf 447}, 3486 (2015), 1412.6620.

\bibitem{1972A&A....20...55D}
J.~{de La Noe} and A.~{Boischot},
\newblock \aap {\bf 20}, 55 (1972).

\bibitem{2013A&A...556A...2V}
M.~P. {van Haarlem} {\em et~al.},
\newblock \aap {\bf 556}, A2 (2013), 1305.3550.

\bibitem{2016ExA....42...11K}
A.~{Konovalenko} {\em et~al.},
\newblock Experimental Astronomy {\bf 42}, 11 (2016).

\bibitem{1979SoPh...62..145A}
E.~P. {Abranin} {\em et~al.},
\newblock \solphys {\bf 62}, 145 (1979).

\bibitem{2010AIPC.1206..445M}
V.~N. {Melnik} {\em et~al.},
\newblock {Type IIIb bursts and their fine structure in frequency band 18-30
  MHz},
\newblock in {\em American Institute of Physics Conference Series}, edited by
  S.~K. {Chakrabarti}, A.~I. {Zhuk}, and G.~S. {Bisnovatyi-Kogan}, , American
  Institute of Physics Conference Series Vol. 1206, pp. 445--449, 2010.

\bibitem{1961ApJ...133..983N}
G.~{Newkirk}, Jr.,
\newblock \apj {\bf 133}, 983 (1961).

\bibitem{2015A&A...580A.137K}
V.~{Krupar} {\em et~al.},
\newblock \aap {\bf 580}, A137 (2015), 1507.06874.

\bibitem{2001A&A...375..629K}
E.~P. {Kontar},
\newblock \aap {\bf 375}, 629 (2001).

\bibitem{2013SoPh..285..217R}
H.~A.~S. {Reid} and E.~P. {Kontar},
\newblock \solphys {\bf 285}, 217 (2013), 1209.5347.

\bibitem{1989SoPh..120..369M}
D.~B. {Melrose},
\newblock \solphys {\bf 120}, 369 (1989).

\bibitem{2014A&A...568A..67M}
D.~E. {Morosan} {\em et~al.},
\newblock \aap {\bf 568}, A67 (2014), 1407.4385.

\bibitem{2015A&A...580A..65M}
D.~E. {Morosan} {\em et~al.},
\newblock \aap {\bf 580}, A65 (2015), 1507.07496.

\bibitem{2015A&A...579A..69O}
M.~{Obrocka}, B.~{Stappers}, and P.~{Wilkinson},
\newblock \aap {\bf 579}, A69 (2015), 1502.06825.

\bibitem{2017A&A...606A.141R}
H.~A.~S. {Reid} and E.~P. {Kontar},
\newblock \aap {\bf 606}, A141 (2017), 1706.07410.

\bibitem{1997PASP..109..166C}
J.~J. {Condon},
\newblock \pasp {\bf 109}, 166 (1997).

\bibitem{1977A&A....61..777B}
J.~L. {Bougeret} and J.~L. {Steinberg},
\newblock \aap {\bf 61}, 777 (1977).

\bibitem{2016JGRA..12111605S}
K.~{Sasikumar Raja} {\em et~al.},
\newblock Journal of Geophysical Research (Space Physics) {\bf 121}, 11 (2016),
  1611.04282.

\bibitem{2012SoPh..275...17L}
J.~R. {Lemen} {\em et~al.},
\newblock \solphys {\bf 275}, 17 (2012).

\bibitem{1974A&AS...15..417H}
J.~A. {H{\"o}gbom},
\newblock \aaps {\bf 15}, 417 (1974).

\bibitem{1995ASPC...75..318K}
U.~{Klein} and K.-H. {Mack},
\newblock {Cleaning of Multi-beam Data},
\newblock in {\em Multi-Feed Systems for Radio Telescopes}, edited by D.~T.
  {Emerson} and J.~M. {Payne}, , Astronomical Society of the Pacific Conference
  Series Vol.~75, p. 318, 1995.

\end{thebibliography}

\end{document}